\documentclass[11pt,twoside]{article}
\usepackage{asp2014}

\aspSuppressVolSlug
\resetcounters

\begin{document}

\title{The KM3NeT Open Science System}

\author{Jutta~Schnabel$^1$, Tamas~Gal$^1$, and Zineb~Aly$^2$}
\affil{$^1$Erlangen Centre for Astroparticle Physics (ECAP), Erlangen, Germany; \email{jutta.schnabel@fau.de, opendata@km3net.de}}
\affil{$^2$ Aix Marseille Univ, CNRS/IN2P3, CPPM, Marseille, France}

\paperauthor{Jutta~Schnabel}{jutta.schnabel@fau.de}{0000-0003-1233-7738}{Erlangen Centre for Astroparticle Physics (ECAP)}{Friedrich-Alexander-Universit\"at}{Erlangen-N\"urnberg}{}{91058}{Germany}

\paperauthor{Tamas~Gal}{tamas.gal@fau.de}{0000-0001-7821-8673}{Erlangen Centre for Astroparticle Physics (ECAP)}{Friedrich-Alexander-Universit\"at}{Erlangen-N\"urnberg}{}{91058}{Germany}

\paperauthor{Zineb~Aly}{zaly@km3net.de}{0000-0002-5593-2580}{Aix Marseille University, CNRS-IN2P3, CPPM}{Physics and Sciences of Matter}{Marseille}{}{13000}{France}





\begin{abstract}

The KM3NeT neutrino detectors are currently under construction at two locations in the Mediterranean Sea, aiming to detect the Cherenkov light generated by high-energy relativistic charged particles in sea water. The KM3NeT collaboration will produce scientific data valuable both for the astrophysics and neutrino physics communities as well as for the Earth and Sea science community. An Open Science Portal and infrastructure are under development to provide public access to open KM3NeT data, software and services. In this contribution, the current architecture, interfaces and usage examples are presented.

\end{abstract}

\section{Introduction}

The KM3NeT collaboration has committed to support open science in its research. In order to facilitate the sharing of FAIR data, the demonstrator of an open science system has been developed by the collaboration. It includes not only a platform to provide science-ready data, but also offers open software and tutorials and integrates the offered products with widely-used repositories and platforms. For this and further developments, test data sets and example analysis workflows are crucial to shape data format and access options according to the needs of the scientists. Therefore, example use cases have been provided to provide KM3NeT research results to the fields of astrophysics, neutrino physics and accross domain boundaries. 

\section{The KM3NeT detectors}

The Cubic Kilometre Neutrino Telescope, KM3NeT, is a European neutrino research infrastructure located in the Mediterranean Sea. KM3NeT operates water-Cherenkov detectors from two locations: Toulon, in France, hosts the Oscillation Research with Cosmics in the Abyss (ORCA) detector at a depth of 2475m, while Portopalo di Capo Passero, Sicily, in Italy, hosts the Astroparticle Research with Cosmics in the Abyss (ARCA) detector at a depth of 3400m.

ARCA is characterized by wider spacing of a 3D array of Digital Optical Modules, DOMs, at a spacing optimised for the detection of high-energy cosmic neutrinos in the TeV to PeV energy range. Its forseen 230 strings are 650m long, spaced 90m apart. On the other hand, ORCA is more compact to optimise the detection of atmospheric neutrinos in the GeV range. ORCA will consist of 115 strings in a 20m triangular grid, with a 9m vertical spacing between the DOMs. Currently, there are 6 Detection Units, DUs, operational for ORCA and one DU for ARCA.

At the shore stations of both ORCA and ARCA, computer farms perform the first data filtering and event triggering to search for the signal of neutrinos, prior to streaming data to central KM3NeT data centres for storage and further analysis by KM3NeT scientists. 

\subsection{Scientific goal}

Beyond the detection of astrophysical neutrinos and neutrino oscillation studies as primary objectives, KM3NeT detectors have a wide range of capabilities. In fact, KM3NeT data can be used for Supernova bursts studies, searches for physics beyond the standard model, acoustic neutrino detection and for Earth and Sea science applications such as bioluminescence activity, whale songs or dolphin clicks. More details on the motivations behind these objectives can be found in \citep{Adri_n_Mart_nez_2016}.

\subsection{Data sets and data formats}
The basic data structure produced from the neutrino detectors is an event, which can be reduced to an array of values per event including particle arrival time, direction, energy, classification, processing parameters as well as uncertainties on derived quantites. However, as the format definition for public data sets are guided by community standards or are developed in exchange with the prospective users, the format varies depending on the scientific context. 

For astrophysics data, the available standards set by the IVOA are applied, and data from a point-source search with the ANTARES detector \citep{ant2007} is offered as testdata set through a KM3NeT-hosted server with the DaCHS software \citep{Dachs}. This allows to perform e.g. a Simple Cone Search, or retrieve the data set as VOTable. For particle events with high significance for multi-messenger analysis, the data is offered as VOEvent through the KM3NeT alert system, see \citep{P4-127_adassxxx}.

However, for event data to be used beyond the Virtual Observatory environment, the KM3NeT Open Data Centre (ODC, has been set up, a platform providing event tables as HDF5 or FITS files and the associated meta data through a REST-API. As a test data sample, all events detected by the ORCA detector within one week with 4 DUs were processed and made available as reference point for further format developments.

In the ODC, also additional high-level science data like event distributions from simulations is offered to support the interpretation of the event tables. It also serves as interface to acoustic data samples measured with a hydrophone hosted in the ORCA detector as an example for environmental data useable for sea science.

\subsection{Analysis examples}

Analysis examples employing a python interface software and test data samples are served in the form of Jupyter notebooks. As example for a search for neutrino point sources, the ANTARES neutrino sample is combined with supplementary services providing a background event estimate and the detector acceptance.

For the ORCA test data set, four analysis examples are offered. In the first three examples, the continuity of data-taking and the event distribution in local coordinates is analysed, the use of the reconstructed direction and quality parameters is used to select events with increased probability to be of astrophyical origin, and the event sample is converted to galactic coordinates. In the fourth example, the galactic coordinates are used to search for events coincident in space and time with a Gravitational Wave event provided by the Gravitational-wave Candidate Event Database of LIGO.

In order to promote a quick understanding of how to analyse KM3NeT data, an education portal offering self-paced online courses has been set up. It offers information on KM3NeT science and a detailed introduction to the use of software analysis tools, facilitating the integration of KM3NeT data in user-defined analyses.

\section{Data access and sharing}

\subsection{The KM3NeT software landscape}

KM3NeT uses a self-hosted GitLab instance as the main platform to develop
software. The continuous integration (CI) feature is a powerful automation tool and is
utilised to generate consistently up-to-date test reports, documentation and
software releases for a large variety of software and firmware including
scientific analyses and even documents like research papers. The CI utilises Docker containers, and KM3NeT adopts this solution for operating-system-level virtualisation and provides Docker and Singularity containers of all major software to facilitate easy software use in versatile contexts.

KM3NeT develops open-source software for accessing and working with data
taken by the detector, produced in simulations or in other analysis pipelines
e.g. event reconstructions. All data produced by the detector and consecutive
processing pipelines which are presented to the public are readable using
Python-based tools, which are released each time a new version is tagged. The choice of Python is mainly motivated by its ease of use and massively increased popularity in a wide range of scientific research areas during the past years.  Each release is
uploaded to the Python Package Index (PyPI) and
can be installed via the pip-package installer. As examples, the km3pipe and km3io packages allow handling of various data formats used in the KM3NeT experiment and include a provenance demonstrator for high-level data processing. Public data from the ODC can be accessed using the openkm3 package. It interlinks with the ODC REST-API and allows to query metadata of the resources and collections. 

All projects which are declared as open source and are hosted on the KM3NeT
GitLab server are visible to the public. However, only KM3NeT members are able to log in and create content on the GitLab server. To enable the interaction with users outside the KM3NeT collaboration, open source software is mirrored to GitHub. The GitHub organisation ``KM3NeT''
has been created to collect all the related projects. It is also planned to link KM3NeT software to the ESCAPE Open Science and Software Repository.

\subsection{Interlinking to repositories and registries}

Registering the KM3NeT open science products with well-used platforms and assigning global identifiers such as DOIs is a key to findability of the data. In the VO, the KM3NeT server is registered as a registry of resources, making KM3NeT data fully findable within the Virtual Observatory, and each resource being identifiable through the naming of the individual endpoint of the service within the registry.

Although the VO does support the integration of a DOI in their resource metadata, it does not provide an authority to assign DOIs. To this end, Zenodo was chosen as well-established data repository in the physics community to obtain a DOI through mirroring the data to the repository. For this demonstrator, the event sample from the KM3NeT ORCA use case will be registered with Zenodo.

\section{Linking to the EOSC}

The current setup foremost serves as a starting point for further developments to improve data sharing with observatories and scientists. Being involved in the ESCAPE project, the KM3NeT collaboration pursues a deepened integration and technology sharing of the open science system with the wider scientfic community in both astroparticle and particle physics. 

This includes efforts to enhance the understanding of how to integrate neutrino data in the Virtual Observatory environment to increase its use in multimessenger analyses, especially regarding the necessity to include theoretical predictions for neutrino measurements with the detectors for neutrino source models. Sharing of e.g. detector sensitivity estimates for the full KM3NeT detector is aimed for not only in the astrophysics context, but also for the area of particle physics, and the specifications to make also this simulation-derived data FAIR are currently investigated.

The ESCAPE project also offers the platform to further pursue standardiziation of data processing, metadata assignment and common format development. Regarding data processing, a sound provenance scheme for the complex event processing workflow is developed drawing on the experience of and in exchange with ESCAPE partners, and the standardization of data formats beyond the VO as well as common simulation of airshowers especially in cooperation with the CTA Observatory is pursued. The development of the EOSC services will also help to pick up these solutions for data storage, sharing, software development and workflow integration in the KM3NeT context.

\section{Summary}
With the KM3NeT Open Data Centre, Virtual Observatory server, Gitlab software development platform and Open Science and Education Portals in place, the KM3NeT collaboration has developed an architecture covering all major requirements for open science and data sharing and is dedicated to improving open science by building on the architecture presented here. 

All relevant links can be found at \href{http://openscience.km3net.de}{http://openscience.km3net.de}.

\bibliography{P9-226}


\end{document}